x
\documentclass[aps,preprint,epsfig,rotate]{revtex4}
\usepackage{graphicx}
\usepackage{bm}
\usepackage{epsfig}

\input epsf
\begin{document}
\title{Highly accurate calculations of the rotationally excited bound
       states in three-body systems}

 \author{Alexei M. Frolov}
 \email[E--mail address: ]{afrolov@uwo.ca}

 \author{David M. Wardlaw}
 \email[E--mail address: ]{dwardlaw@uwo.ca}

\affiliation{Department of Chemistry\\
 University of Western Ontario, London, Ontario N6H 5B7, Canada}

\date{\today}

\begin{abstract}

An effective optimization strategy has been developed to construct highly
accurate bound state wave functions in various three-body systems. Our
procedure appears to be very effective for computations of weakly bound
states and various excited states, including rotationally excited states,
i.e. states with $L \ge 1$. The efficiency of our procedure is illustrated
by computations of the excited $P^{*}(L = 1)-$states in the $dd\mu, dt\mu$
and $tt\mu$ muonic molecular ions, $P(L = 1)-$states in the non-symmetric
$pd\mu, pt\mu$ and $dt\mu$ ions and $2^{1}P(L = 1)-$ and $2^{3}P(L =
1)-$states in He atom(s).

\end{abstract}

\maketitle

\newpage

In this study we develop a new optimization strategy which is used to
construct extremely accurate variational wave functions in arbitrary
three-body systems. As follows from the results of numerical computations
this procedure is a very effective method to determine highly accurate,
bound state wave functions in various three-body atoms, ions, muonic ions,
etc. It can be applied not only to ground states, but, in principle, to
arbitrary excited states in three-body systems, including very weakly bound
states, rotationally and vibrationally excited states, Rydberg states in
atoms, etc. Our computational goal in this work is to determine the highly
accurate solutions of the corresponding Schr\"{o}dinger equation for the
bound state spectra $H \Psi = E \Psi$, where $E < 0$, and $H$ is the
non-relativistic Hamiltonian of an arbitrary three-body system. In
particular, for Coulomb three-body systems the Hamiltonain $H$ takes the
form
\begin{equation}
 H = - \frac{\hbar^2}{2 m_a} \Bigl(\frac{m_a}{m_1} \nabla_1^2 +
 \frac{m_a}{m_2} \nabla_2^2 + \frac{m_a}{m_3} \nabla_3^2\Bigr) +
 \frac{q_3 q_2 e^2}{r_{32}} + \frac{q_3 q_1 e^2}{r_{31}} + \frac{q_2
 q_1 e^2}{r_{21}}
\end{equation}
where $m_1, m_2, m_3$ and $q_1, q_2, q_3$ are the particle masses and
charges, respectively. Also, in this equation $\hbar = \frac{h}{2 \pi}$ is
the reduced Planck constant, $e$ is the electron's electric charge and $m_a$
is the mass of some elementary particle. Below, we shall use the following
units: $\hbar = 1, e = -1$ and $m_a = \min(m_1, m_2, m_3) = 1$. Such a
choice of units corresponds to the atomic units in the case of atoms/ions,
where $m_a = m_e$, and to the muon-atomic units in the case of muonic
atoms/ions and muon-molecular ions, where $m_a = m_{\mu}$. We shall not
assume $a$ $priori$ that some particle masses are infinite.

In this work, all solutions of the Schr\"{o}dinger equation are approximated
with the use of a finite set of exponents which explicitly depend upon three
relative coordinates $r_{32}, r_{31}$ and $r_{21}$. In the case of bound
states with angular momentum $L$ ($L \ge 1$) this exponential ansatz is
written in the form \cite{Fro1}
\begin{equation}
 \Psi = \frac{1}{2} (1 + \kappa \hat{P}_{21}) \sum_{i=1}^{N}
 \sum_{\ell_1=0}^{L} C_{i}(\ell_1) {\cal Y}^{(\ell_1,\ell_2)}_{L0}({\bf
 n}_{32}, {\bf n}_{31}) \exp(-\alpha_i r_{32} - \beta_i r_{31} - \gamma_i
 r_{21}) \label{exp1}
\end{equation}
where $r_{ij} = r_{ji}$ are the three relative coordinates, ${\bf n}_{3i} =
\frac{{\bf r}_{3i}}{r_{3i}} = -{\bf n}_{i3}$ and $\ell_1 + \ell_2 = L$ (or
in some cases $\ell_1 + \ell_2 = L + 1$, see below). The operator
$\hat{P}_{21}$ in this equation is the permutation of the two identical (1
and 2) particles, e.g., electrons, muons, nuclei. The parameter $\kappa$
equals zero for non-symmetric three-body systems, while for symmetric
three-body systems we have $\kappa = (-1)^L$. The coefficients $C_i$ in
Eq.(\ref{exp1}) are the linear variational parameters of the method, which
are determined by solving the corresponding Schr\"{o}dinger equation. The
parameters $\alpha_i, \beta_i, \gamma_i$ ($i = 1, \ldots, N$) in
Eq.(\ref{exp1}) are the non-linear parameters of the exponential expansion.
Also, in Eq.(\ref{exp1}) the functions ${\cal Y}^{(\ell_1,\ell_2)}_{LM}({\bf
n}_1, {\bf n}_2)$ are the bi-polar harmonics (see, e.g., \cite{Varsh}). The
quantity $L$ is the total angular momentum of the considered bound state
in the three-body system, while $M$ is the magnetic quantum number. In the
absence of an external magnetic field, all considered bound state
properties, including the total energies, cannot depend upon the numerical
value of $M$. Therefore, without loss of generality we will assume that $M =
0$ in Eq.(\ref{exp1}). The individual angular momenta $\ell_1$ and $\ell_2$
in Eq.(\ref{exp1}) are integer, non-negative numbers and their sum equals
$L$, since in this work we consider only bound states with natural
(spatial) parity.

The variational expansion Eq.(\ref{exp1}) and its various modifications
(see, e.g., \cite{Fro1}, \cite{BaFr}) are extensively used in bound state
computations of many different three-body systems. Note that the first
numerical computations to use the exponential variational expansion,
Eq.(\ref{exp1}), were performed in 1968 \cite{DK} by Delves and Kalotas for
the ground $S(L = 0)-$state of the $pp\mu$ muonic molecular ion. Since that
1968 study \cite{DK} the overall efficiency of the exponential variational
expansion Eq.(\ref{exp1}) has been increased substantially. Currently, this
expansion \cite{Fro1}, \cite{BaFr} is one of the most effective methods
specifically designed for high precision variational calculations of the
bound state spectra in arbitrary three-body systems. The very high
efficiency of this expansion arises from the advanced optimization strategy
used to chose the non-linear parameters in Eq.(\ref{exp1}). In general, such
a strategy includes the two following stages: (1) construction of highly
accurate ``short-term'' cluster wave function (so-called booster wave
function) which includes $N_0 \approx$ 400 - 600 exponential basis
functions, and (2) optimization of the remaining part of the total wave
function which includes $N - N_0 \ge 3000$ exponential basis functions.
Optimization of the non-linear parameters in the short-term wave function is
performed very carefully and accurately. Analogous optimization of 28
non-linear parameters at the second step of our procedure \cite{Fro1},
\cite{Fro6} can be accurate only for relatively small dimensions $N \approx
800 - 1500$. At larger dimensions any careful optimization of non-linear
parameters in the total wave function takes a very long time. In \cite{Fro6}
we have developed an approximate optimization method for these parameters
which was based on a separation of these 28 non-linear parameters into three
groups: fast, intermediate and slow parameters depending on the rate of
convergence of the optimization. The optimization was performed only for the
fast and intermediate non-linear parameters. In general, this procedure
works very well. This strategy was successfully used in applications to the
Ps$^-$ and H$^-$ ions, He atoms (ground $1^1S-$state and excited
$2^3S-$state) and to many other three-body systems.

In this study our main goal is to consider the rotationally excited bound
states, i.e. states with $L \ge 1$ in three-body systems with arbitrary
masses of three particles. To compute these states to high numerical
accuracy below we have developed a different idea of approximate
optimization of the non-linear parameters at large and very large
dimensions. This idea can be illustrated by considering only one non-linear
parameter (of 28 such parameters \cite{Fro1}), e.g., the $A_2$ parameter
(all notations used here and below were defined in \cite{Fro1}). In general,
the optimal value of this parameter varies when the total number of basis
functions increases. In other words, we have a $A_2(N)$ function, where $N =
N_0 + (N - N_0)$ is the total number of basis functions in Eq.(\ref{exp1}).
Here we consider the situation when the booster function with $N_0$ terms
has been constructed already, and it does not change at the following steps.
Suppose, for simplicity, that $N_0 = 400$ and at the second stage we need to
perform optimization of 28 non-linear parameters by using wave functions
with $N$ = 800, 1000, 1200 and 1400 terms. After optimizations at each
dimension we obtain four `optimal' values for the $A_2$ parameter, i.e.
$A_2(800), A_2(1000), A_2(1200)$ and $A_2(1400)$. As follows from the
results of numerical calculations, the overall variations of the $A_2(N)$
values for $N$ = 800, 1000, 1200 and 1400 are typically relatively small.
Therefore, we can determine the approximate limit of the $A_2(N)$ function
for $N \rightarrow \infty$. The projected value $A_2$, or $A_2(\infty)$, can
be used in actual calculations with large and very large number of basis
functions. The same procedure is applied to the 27 remaining non-linear
parameters used in our method as described in \cite{Fro1}, \cite{Fro6}. This
gives us the approximate optimal values of all 28 non-linear parameters
$A_1(\infty), A_2(\infty), B_1(\infty), B_2(\infty), \ldots$ needed at the
second stage of our method. The overall accuracy achieved at the second
stage of our procedure is even better than the approximate procedure in
Ref.\cite{Fro6} can provide. Note that in this method we do not neglect any
of the non-linear parameters (compare with the ignorance of `slow'
non-linear parameters in \cite{Fro6}).

To demonstrate the efficiency of our new procedure for construction of
highly accurate wave functions, let us consider its application to the
weakly bound $P^{*}(L = 1)-$states in the $dd\mu, dt\mu$ and $tt\mu$ muonic
molecular ions. Here the notation $L$ designates the total angular momentum
of the three-body system, while the asterisk `*' means that the wave
function of this state equals zero for some configuration in relative
coordinates $r_{32}, r_{31}$ and $r_{21}$. This classification scheme
originated from the one-electron hydrogen atom. In the alternative system of
$(L, \nu)$-notations this state is designated as the (1,1)-state, where $L =
1$ means `rotationally' excited state and $\nu = 1$ means `vibrationally'
excited state. The alternative classification system for bound states in
three-body systems originated from the two-center H$^{+}_2$ ion which is a
pure adiabatic (or Born-Oppenheimer \cite{BoOp}) system. Despite some
differences in these classification schemes one easily finds a uniform
correspondence between the explicit notations used to designate the same
bound states in three-body systems. It is well known that the bound
$P^{*}(L = 1)-$states in the $dd\mu$ and $dt\mu$ ions are extremely weakly
bound. In general, a state can be considered as weakly bound if its binding
energy is $\le$ 1 \% of its total energy \cite{BiFr}, \cite{Fro92}; for the
$P^{*}(L = 1)-$state in the $dt\mu$ ion the binding energy is less than
0.01 \% of its total energy. Highly accurate calculations of such states are
very difficult to perform.

Note that each of the $dd\mu, dt\mu$ and $tt\mu$ ions is a Coulomb
three-body system with unit charges. The general theory of such systems was
developed twenty years ago (see, e.g., \cite{BiFr}, \cite{Fro92} and
references therein). In those works it was shown that the type of bound
state spectra in an arbitrary $a^{+} b^{+} \mu^{-}$ system is determined
substantially by the lightest positive ion, e.g., by the $a^{+}$ ion, if
$m_a \le m_b$. Briefly, this means that the binding energies of the $a^{+}
b^{+} \mu^{-}$ and $a^{+} a^{+} \mu^{-}$ ions are always very close to each
other, while their total energies can be substantially different. Moreover,
as follows from \cite{Fro92}, the binding energy of the $a^{+} b^{+}
\mu^{-}$ ion is always smaller than the binding energy of the $a^{+} a^{+}
\mu^{-}$ ion, if $m_a \le m_b$. In application to the muonic molecular ions
mentioned above this means that the binding energies of the $P^{*}(L =
1)-$states in the $dd\mu$ and $dt\mu$ ions are approximately equal to each
other. This is a very interesting situation since the total energies of the
$P^{*}(L = 1)-$states in the $dd\mu$ and $dt\mu$ ions are substantially
different from each other (the actual difference is $\approx$ 60 $eV$).
Furthermore, since the $P^{*}(L = 1)-$state in the $dd\mu$ ion is weakly
bound, then the analogous state in the $dt\mu$ ion is even more weakly
bound. On the other hand, since $m_{t} > m_{d}$, then it is clear that the
energy spectrum of the $tt\mu$ ion is completely different from those of the
$dd\mu$ and $dt\mu$ ions. In particular, the bound state spectra of the
$tt\mu$ ion contains six (not five!) bound states and the $P^{*}(L =
1)-$state of the $tt\mu$ ion is not weakly bound.

The results of numerical calculations of the $P^{*}(L = 1)-$states in the
$dd\mu, dt\mu$ and $tt\mu$ muonic molecular ions can be found in Table I. In
our calculations we have used the following values of nuclear masses
\cite{COD}, \cite{CRC}:
\begin{eqnarray}
 m_{\mu} = 206.268262 m_e \; \; \; , \; \; \; m_p = 1836.152701 m_e \\
 m_{d} = 3670.483014 m_e \; \; \; , \; \; \; m_t = 5496.92158 m_e \nonumber
\end{eqnarray}
where $m_e$ designates the electron mass. All computations are performed in
muon-atomic units, where $m_{\mu} = 1, \hbar = 1$ and $e = 1$. Note that our
highly accurate computations in this study are performed with the use of 84
- 104 decimal digits per computer word \cite{Bail1}, \cite{Bail2}. The total
energies are determined to the accuracy $\approx 1 \cdot 10^{-20} - 1 \cdot
10^{-23}$ $m.a.u.$ (or $a.u.$ for atoms). For the purposes of reporting the
results of our calculations in this paper, it was assumed that all particle
masses and corresponding conversion factors (e.g., the factor $Ry$ below)
are exact. Such assumptions are always made in papers on highly accurate
computations in few-body systems (see, e.g., \cite{BaFr}, \cite{Drak}). The
known experimental uncertainties in particle masses and conversion factors
can be taken into account at the last step of calculations, when the most
accurate computation would simply be repeated with the use of particle
masses and conversion factors of known or chosen acccuracy and the resulting
energies reported to to an accuracy commensurate with the accuracy of these
masses and conversion factors. A primary motivation for, and advantage of,
calculating wave functions of very high accuracy is that relatively small
energy differences between states in a given system or between systems can
be reliably detected and and calculated, after which the requisite
conversion factor can be applied and account taken of its accuracy.

As follows from Table I the total energies obtained for these states are
significantly more accurate than analogous energies determined in earlier
studies. The $P^{*}(L = 1)-$states in the $dd\mu$ and $dt\mu$ ions are of
great interest in some applications (see, e.g., \cite{BaFr}, \cite{BD} and
references therein). The binding $\varepsilon$ and total $E$ energies of the
$P^{*}(L = 1)-$states in the $dd\mu$ and $dt\mu$ ions determined from the
results shown in Table I are:
\begin{eqnarray}
 E(dd\mu) = -0.473 686 733 842 727 0 \pm 5 \cdot 10^{-16} m.a.u \; \; \; ,
 \varepsilon(dd\mu) = -1.974 988 087 997(3) eV \\
 E(dt\mu) = -0.481 991 529 973 85 \pm 5 \cdot 10^{-14} m.a.u \; \; \; ,
 \varepsilon(dt\mu) =  -0.660 338 686 5(3) \; \; eV \nonumber
\end{eqnarray}
where the conversion factor $Ry = 27.2113961 \Bigl( \frac{m_{\mu}}{m_{e}}
\Bigr)$ was used. The binding energies are related to the total energies as
$\varepsilon(dd\mu) = [E(dd\mu) - E(d\mu)] Ry$ and $\varepsilon(dt\mu) =
[E(dt\mu) - E(t\mu)] Ry$, where $E(d\mu) = -\frac12 \frac{m_{d}
m_{\mu}}{m_{d} + m_{\mu}}$ and $E(t\mu) = -\frac12 \frac{m_{t}
m_{\mu}}{m_{t} + m_{\mu}}$ are the total energies of the ground states in
the $d\mu$ and $t\mu$ muonic atoms, respectively. By performing such a
re-calculation from the total energy (expressed in $m.a.u.$) to the binding
energy (expressed in $eV$) we assume that the factor $Ry$ is exact.

Table II contains results of variational computations of the three $P(L =
1)-$states in the non-symmetric $pd\mu, pt\mu$ and $dt\mu$ muonic molecular
ions. These bound states are traditionally difficult for highly accurate
computations. Nevertheless, by using our approach described above we have
determined their total energies to very high accuracy, significantly
exceeding numerical accuracy achieved in earlier calculations. The total
energies of three bound $P(L = 1)-$states in the non-symmetric muonic
molecular ions $pd\mu, pt\mu$ and $dt\mu$ are now known with numerical
uncertainties less than $1 \cdot 10^{-15}$ $m.a.u.$ which is sufficient for
all practically important problems which include these ions.

To confirm the efficiency of our method we have also calculated the
rotationally excited $2^1P(L = 1)-$ and $2^3P(L = 1)-$states in the
${}^{\infty}$He, ${}^{3}$He and ${}^{4}$He helium atoms. The total energies
of these states in the ${}^{\infty}$He, ${}^{3}$He and ${}^{4}$He helium
atoms have been computed in a number of earlier studies (see, e.g.,
\cite{Drak}, \cite{Acad}, \cite{Fro03}). Our values of the total energies
(in atomic units $\hbar = 1, m_e = 1, e = 1$) of these states can be found
in Table III. In calculations of the ${}^3$He and ${}^4$He atoms we have
used the following values of nuclear masses \cite{COD}, \cite{CRC}
\begin{equation}
 M_{{}^{3}He^{2+}} = 5495.8852 m_e \; \; , \; \; M_{{}^{4}He^{2+}} =
 7294.2996 m_e
\end{equation}
Note that our calculations of the $2^1P(L = 1)-$ and $2^3P(L = 1)$-states of
the helium atoms have been performed in atomic units. As follows from Table
III our total energies determined for these two states in the helium atoms
are substantially more accurate than the total energies known for these
states from the literature.

In conclusion, the results of this study indicate that the exponential
variational expansion in relative coordinates, Eq.(\ref{exp1}), appears to
be the most appropriate expansion for high precision variational
calculations of the bound state spectra in non-relativistic three-body
systems. For an arbitrary Coulomb three-body system (with arbitrary particle
masses $m_1, m_2, m_3$) the approach based on Eq.(\ref{exp1}) allows one to
obtain extremely accurate (i.e. essentially exact) numerical solutions for
the ground and all excited states, including weakly bound states and
Rydberg states in two-electron atoms \cite{BS}, \cite{Sob}. The
generalization of Eq.(\ref{exp1}) to the case of adiabatic (or two-center)
systems can be found in \cite{Fro02}. To provide a very high accuracy for
the bound states in adiabatic systems, e.g., for H$^{+}_2$ ion, the
non-linear parameters $\alpha_i, \beta_i, \gamma_i$ in Eq.(\ref{exp1}) must
be complex, i.e. they must have non-zero imaginary parts. It is also
important to note that numerical accuracy of such solutions can be made
arbitrarily high, e.g., by using more advanced optimization strategies for
actual non-nonlinear parameters. The highly accurate wave functions
obtained in this study are needed to determine many bound state properties
of muonic molecular ions to the accuracy which will be sufficient for
current experimental capabilities. In particular, the knowledge of highly
accurate variational wave functions is extremely important for weakly-bound
$P^{*}(L = 1)-$states in the $dt\mu$ and $dd\mu$ muonic molecular ions. In
addition, there is a need for the prediction of the lowest order
relativistic, QED, and mass corrections to the total non-relativistic energy
in arbitrary three-body systems and the computational results for such
corrections depend sensitively on the accuracy of the ``zeroth-order''
wave function.

\begin{center}
    {\bf Acknowledgements}
\end{center}

It is a pleasure to thank David H. Bailey (Lawrence Berkeley National
Laboratory, Berkeley, California) for his valuable help and discussions and
the University of Western Ontario for financial support.

\newpage
%
%
   \begin{table}[tbp]
    \caption{The total energies $E$ of the excited $P^{*}(L = 1)-$states
             (or (1,1)-states) in the $dd\mu, tt\mu$ and $dt\mu$ muonic
             molecular ions in muon-atomic units ($m_{\mu} = 1, \hbar = 1,
             e = 1$). $N$ designates the number of basis functions used in
             Eq.(2).}
     \begin{center}
     \scalebox{0.95}{%
     \begin{tabular}{llll}
        \hline\hline
$N$ & $E(dd\mu; (1,1)-$state) & $E(tt\mu; (1,1)-$state) & $E(dt\mu; (1,1)-$state) \\
       \hline
 3300 & -0.473 686 733 842 725 810 43 & -0.489 908 667 504 942 696 78 & -0.481 991 529 973 471 68 \\

 3500 & -0.473 686 733 842 725 998 08 & -0.489 908 667 504 942 819 82 & -0.481 991 529 973 590 96 \\

 3700 & -0.473 686 733 842 726 125 12 & -0.489 908 667 504 942 935 53 & -0.481 991 529 973 677 08 \\

 3800 & -0.473 686 733 842 726 218 22 & -0.489 908 667 504 942 980 54 & -0.481 991 529 973 697 42 \\

 3840 & -0.473 686 733 842 726 242 81 & -0.489 908 667 504 942 990 80 & -0.481 991 529 973 715 97 \\
              \hline
 $A^a$ & -0.473 686 733 842 727 0(5) & -0.489 908 667 504 943 30(8) & -0.481 991 529 973 85(5) \\

 $E^b$ & -0.473 686 733 842 720 3 \cite{BaFr} & -0.489 908 667 504 93 \cite{Fro1} &
-0.481 991 529 971 713 \cite{BaFr} \\
   \hline \hline
  \end{tabular}}
  \end{center}
${}^a$The expected energies with estimated uncertainties. \\
${}^b$The best variational energies known from earlier
calculations. \\
  \end{table}
%
%
   \begin{table}[tbp]
    \caption{The total energies $E$ of the $P(L = 1)-$states
             (or (1,0)-states) in the $pd\mu, pt\mu$ and $dt\mu$ muonic
             molecular ions in muon-atomic units ($m_{\mu} = 1, \hbar = 1,
             e = 1$). $N$ designates the number of basis functions used in
             Eq.(2).}
     \begin{center}
     \scalebox{0.95}{%
     \begin{tabular}{llll}
        \hline\hline
$N$ & $E(pd\mu; (1,0)-$state) & $E(pt\mu; (1,0)-$state) & $E(dt\mu; (1,0)-$state)  \\
     \hline
 3300 & -0.490 664 169 479 316 12 & -0.499 492 029 991 533 40 & -0.523 191 456 315 954 69 \\

 3500 & -0.490 664 169 479 320 24 & -0.499 492 029 991 534 83 & -0.523 191 456 315 955 64 \\

 3700 & -0.490 664 169 479 323 29 & -0.499 492 029 991 535 83 & -0.523 191 456 315 956 51 \\

 3840 & -0.490 664 169 479 324 78 & -0.499 492 029 991 536 26 & -0.523 191 456 315 957 14 \\
              \hline
 $A^a$ & -0.490 664 169 479 327(1) & -0.499 492 029 991 539(1) &
     -0.523 191 456 315 960(1) \\

 $E^b$ & -0.490 664 169 479 315 \cite{Fro1} & -0.499 492 029 991 513
 \cite{Fro1} & -0.523 191 456 315 937 14 \cite{Fro1} \\
       \hline\hline
  \end{tabular}}
  \end{center}
${}^a$The expected energies with estimated uncertainties. \\
${}^b$The best variational energies known from earlier
calculations. \\
  \end{table}
%
%
  \begin{table}[tbp]
  \caption{The total energies $E$ (in atomic units $m_e = 1,
            \hbar = 1, e = 1$) for the $2^1P(L = 1)-$ and
            $2^3P(L = 1)-$states in the ${}^{\infty}$He atom.
            $N$ designates the number of basis functions used
            in Eq.(2).}
     \begin{center}
     \scalebox{0.95}{%
     \begin{tabular}{lll}
        \hline\hline
$N$ & $2^1P(L = 1)-$state & $2^3P(L = 1)-$state \\
        \hline
 3500 & -2.123 843 086 498 101 358 895 90 &
        -2.133 164 190 779 283 205 057 11 \\

 3800 & -2.123 843 086 498 101 358 970 11 &
        -2.133 164 190 779 283 205 079 75 \\

 4000 & -2.123 843 086 498 101 359 030 07 &
        -2.133 164 190 779 283 205 092 36 \\

 4200 & -2.123 843 086 498 101 359 074 18 &
        -2.133 164 190 779 283 205 102 51 \\
              \hline
 $A^a$ & -2.123 843 086 498 101 359 20(5) &
             -2.133 164 190 779 283 205 17(5) \\
      \hline
 $E^b$ &
 -2.123 843 086 498 101 360(2) \cite{Fro03} &
 -2.133 164 190 779 283 206(2) \cite{Fro03} \\

 $E^b$ &
 -2.123 843 086 498 093(2) \cite{Drak} &
 -2.133 164 190 779 273(5) \cite{Drak} \\
      \hline
 $E^c({}^3$He) &
 -2.123 448 345 012 547 695 33 &
 -2.132 787 874 710 055 466 03 \\

  $E^c({}^4$He) &
 -2.123 553 590 529 057 856 32 &
 -2.132 880 642 105 551 984 62 \\
       \hline\hline
  \end{tabular}}
  \end{center}
${}^a$The expected energies with estimated uncertainties. \\
${}^b$The best variational energies known from earlier calculations. \\
${}^c$Results obtained for $N = 3800$. \\
  \end{table}
\end{document}